\documentclass[aps,11pt,tightenlines,nofootinbib]{revtex4} 

\usepackage{amsmath,amssymb,amsthm}

\def\paper{Article}

\def\ra{\rightarrow}

\def\C{\mathcal{C}}

\def\R{\mathbb{R}}

\def\RTR{\R_+^{1+1}}
\def\Linf{{L^\infty}}

\def\<{\langle}
\def\>{\rangle}

\def\w#1{\widetilde{#1}}
\def\h#1{\widehat{#1}}

\def\d{\partial}

\def\eps{\varepsilon}
\def\la{\lambda}

\def\O{O}

\def\W#1{W^{(#1)}}

\begin{document}

\title{Late-time attractor for the cubic nonlinear wave equation}

\author{Nikodem Szpak}

\affiliation{Max-Planck-Institut f\"{u}r
Gravitationsphysik, Albert-Einstein-Institut, Golm, Germany}
\date{\today}

\begin{abstract}
  We apply our recently developed scaling technique for obtaining late-time asymptotics to the cubic nonlinear wave equation and explain appearance and approach to the two-parameter attractor found recently by Bizon and Zenginoglu.
\end{abstract}


\maketitle

In \cite{Bizon+Anil} Bizo{\'n} and Zengino\u{g}lu conjecture and present some analytical and numerical evidence that the spherically symmetric cubic nonlinear wave equation in three dimensions
\begin{equation} \label{wave-eq-u3}
  \Box u = u^3,\qquad\qquad (\Box\equiv \d_t^2-\Delta)
\end{equation}
has a universal two-parameter attractor for the late-time asymptotics
\begin{equation} \label{attractor}
  u_{a,b}(t,r) = \frac{\sqrt{2}}{t+a+b[(t+a)^2-r^2]}
\end{equation}
for a big family of initial data, being itself an exact solution of \eqref{wave-eq-u3}.
The rate of approach to the attractor \eqref{attractor} with suitably chosen $a,b$ is $t^{-4}$ for a fixed $r$.
The aim of this \paper\ is to prove this conjecture by deriving a precise late-time asymptotics and comparing it with the form of the attractor \eqref{attractor}.
Recently, in \cite{NS-Scaling}, we have developed a method for obtaining late-time asymptotics from the scaling properties of a given wave equation. It allows us to show that, at least for small initial data, the generic late-time asymptotics of solutions to \eqref{wave-eq-u3} takes the form
\begin{equation} \label{asympt-p=3}
  u(t,r) = \eps u_0(t,r) +
   \frac{A_0}{t^2-r^2} + \frac{A_1 t}{(t^2-r^2)^2} + \frac{A_2 (3t^2+r^2)}{(t^2-r^2)^3} + \dots
\end{equation}
where $A_i$ are given in terms of the initial data.
This expansion coincides with the attractor \eqref{attractor} when $A_0=\sqrt{2}/b$ and $A_1=\sqrt{2}/(2ab+1)$ up to the term of order $t^2(t^2-r^2)^{-3}$. Hence, as long as $A_0, A_1$ are non-zero there exist unique parameters $a,b$ such that the solution $u$ asymptotically approaches the two-parameter attractor \eqref{attractor}. The rate of approach is determined by the failure of \eqref{attractor} to reproduce the third term in the generic asymptotics \eqref{asympt-p=3} which contains a third independent parameter $A_2$.


Such precise asymptotic analysis has only become possible with the recent developments in the perturbation theory. Crucial are the first rigorous works \cite{NS-Tails, NS-PB_Tails} allowing for establishing a link between the late-time asymptotics and the small initial data (see also references therein for earlier non-rigorous but important works, e.g. by Bizo{\'n} \textit{et al}).
They build up on earlier decay estimates for small data by John \cite{John-blowup} and Asakura \cite{Asakura}.
Later, in \cite{NS-Scaling}, an equivalent technique based on scaling has been introduced which simplifies the asymptotic calculations of higher order terms.
In it, the initial value problem for a class of nonlinear wave equations is considered which we restrict  here for the sake of simplicity to
\begin{equation} \label{wave-eq}
  \Box u = u^p
\end{equation}
with integer $p\geq 3$ and small initial data
\begin{equation} \label{init-data}
  u(0,r)=\eps f(r),\qquad \d_t u(0,r)=\eps g(r)
\end{equation}
being smooth functions of compact support 
in three spatial dimensions restricted to spherical symmetry. Then, for small $\eps$, we have
$u\in\C^\infty(\RTR)$.

The main idea of deriving asymptotics from scaling is based on the observation of Lindblad \cite{Lindblad-PhD_CPDE} that in the limit $\eps\ra 0$ the solutions $u_\eps$ of \eqref{wave-eq}-\eqref{init-data} tend, under suitable scaling, to some nontrivial $u_*$ which satisfies a linear wave equation with a distributional source. This equation can be solved exactly.
For small but finite values of $\eps$ the solutions $u_\eps$ are near to $u_*$ in a suitable sense with a uniform error bound such that $u_*$ determines the late-time asymptotics of $u_\eps$.

A straightforward generalization of Theorem 1 of \cite{NS-Scaling} leads to
\begin{equation} \label{u-asympt-n}
\begin{split}
  u(t,r) 
  &= \eps u_0(t,r) +
  \eps^p \sum_{k=0}^n  \frac{B_{p,k}}{r} \left[ \frac{1}{(t-r)^{p+k-2}} - \frac{1}{(t+r)^{p+k-2}} \right]\\
   &+ \O\left(\frac{\eps^p}{\<t+r\>\<t-r\>^{p+n-2}}\right)
  + \O\left(\frac{\eps^{p+\la_0}}{\<t+r\>\<t-r\>^{p-2}}\right)
\end{split}
\end{equation}
for $\eps\ra 0$, $t-r>1/\eps^a$, a given scaling parameter $0<a<p(p-1)/(p+1)$ and any nonnegative integer $n$.
$u_0$ solves the corresponding linear problem \eqref{free-wave}-\eqref{free-wave-init} (see below) and $B_{p,k}$ are determined by the initial data and are defined below.

The error terms mean that the asymptotics holds w.r.t. to weighted-$\Linf$ norms (cf. \cite{NS-Scaling}). Here, they are restricted to the region $t-r>1/\eps^a$ and imply a uniform convergence there, as $\eps\ra 0$.
The first error term describes correction entering at the same nonlinear order (the same power of $\eps$) as the leading terms but having faster decay in time
while the second error term, with $\la_0 := {(p-1)(1-a)+a[(p-1)^2-2]/p}$, stays for corrections with the same decay in time as the leading terms but entering with higher powers of $\eps$.

For $p=3$ and $n=2$ we essentially obtain the asymptotic expansion \eqref{asympt-p=3}. The only point which requires some additional but straightforward work is to show that all expressions appearing in the second error term are actually of the same functional form as those alredy present in the leading asymptotics multiplied by higher powers of $\eps$. Then they don't change the character of the asymptotic expansion but only alter the constants, thus leading to \eqref{asympt-p=3} with $A_k(\eps)=B_{3,k}\eps^3+\O(\eps^4)$.

In \cite[Sec. 5.1.2]{Bizon+Anil} also non-generic solutions with faster late-time decay than this given by \eqref{attractor} have been found numerically. This can be explained by the observation that the initial data can be chosen such that at late times $A_0=0$ or $A_0=A_1=0$, etc., thus leading to faster decay  in \eqref{asympt-p=3}.

\section*{The scaling technique}

Here, we briefly introduce the method of scaling developed in \cite{NS-Scaling} and extend it to calculate higher order terms appearing in \eqref{u-asympt-n}. All technical details and proofs can be found or easily adapted from \cite{NS-Scaling}. The scaling method has an advantage over the standard perturbation theory in generating simpler--to--solve effective equations for the higher order asymptotic terms.

In the first step we solve a corresponding linear equation with removed scale factor $\eps$
\begin{equation} \label{free-wave}
  \Box u_0 = 0,
\end{equation}
\begin{equation} \label{free-wave-init}
  u_0(0,r)=f(r),\qquad \d_t u_0(0,r)=g(r).
\end{equation}
Its solution can be written in the form
\begin{equation} \label{free-u0}
  u_0(t,r)=\frac{h(t-r)-h(t+r)}{r}
\end{equation}
where
\begin{equation} \label{def:h}
  h(x):=-\frac{x}{2} f(x) - \frac{1}{2} \int_x^\infty y g(y) dy
\end{equation}
has compact support (the functions $f(r), g(r)$ have been symmetrically continued to negative $r$).
Next, we subtract the linear solution from the nonlinear one by introducing
\begin{equation}
  w(t,r):=u(t,r) - \eps u_0(t,r)
\end{equation}
which satisfies
\begin{equation}
  \Box w = (w + \eps u_0)^p.
\end{equation}
Now, we scale this function to
\begin{equation} \label{def-Weps}
  W_\eps(t,r):=\eps^{-b} w(\eps^{-a} t, \eps^{-a} r)
\end{equation}
with $b=p+a(p-1)$ and some $a>0$ to be chosen later. It satisfies
\begin{equation} \label{free-W_eps}
\begin{split}
  \Box W_\eps(t,r) &= \eps^{-a} \left[\eps^{(p-1)+a(p-2)} W_\eps(t,r) + \eps^{-a} u_0(\eps^{-a} t, \eps^{-a} r) \right]^p \\
  &= \eps^{-a} \left[ \eps^{-a} u_0(\eps^{-a} t, \eps^{-a} r) \right]^{p} \\
  &+ \sum_{k=1}^p \begin{pmatrix} p\\ k\end{pmatrix} \eps^{-a} \left[ \eps^{-a} u_0(\eps^{-a} t, \eps^{-a} r) \right]^{p-k} \eps^{[(p-1)+a(p-2)]k} W_\eps^k(t,r).
\end{split}
\end{equation}
For this equation we want to consider the limit $\eps\ra 0$.

Let us recall the following fact of the distributional calculus: any smooth function $H(x)$ of compact support can be squeezed to the delta distribution under appropriate scaling as $\eps\ra 0$. The corrections can be written as a sum over derivatives of the delta
\begin{align}
  \eps^{-1} H(\eps^{-1} x) \approx C_0 \delta(x) + \eps C_1 \delta'(x) + \dots
\end{align}
what, in the precise (distributional) sense, means
\begin{align}
  \lim_{\eps\ra 0} \left[\int \frac{1}{\eps^{n+1}} H\left(\frac{x}{\eps}\right) g(x)\ dx
- \sum_{k=0}^{n-1} \frac{1}{\eps^{n-k}} C_k g^{(k)}(0) \right] = C_n g^{(n)}(0)
\end{align}
where $g\in\C_0^\infty$ is a test function, $n$ is any nonnegative integer and $C_k:=\int x^k H(x)\ dx$.

Having this in mind we observe that the first term in \eqref{free-W_eps}, by use of the representation \eqref{free-u0}, will have a distributional limit (in the above sense)
\begin{equation} \label{u0-delta}
  \eps^{-a} \left[ \eps^{-a} u_0(\eps^{-a} t, \eps^{-a} r) \right]^{q} \approx
  C_{q,0} \frac{\delta(t-r)-\delta(t+r)}{r^{q}} +
  \eps^a C_{q,1} \frac{\delta'(t-r)-\delta'(t+r)}{r^{q}} + \dots
\end{equation}
where
\begin{equation} \label{def:Ck}
  C_{q,i} := \int x^i h^q(x) dx.
\end{equation}
The terms $\delta^{(k)}(t+r)$ will further play no role since their support is outside of the region of our interest $t+r>0$.

The sum in \eqref{free-W_eps} will be treated as an error term. The expansion
\begin{align}
  W_\eps = \W0 + \eps^a \W1 + \dots + \eps^{na} \W{n} + \eps^{(n+1)a} \h{W} + \w{W}_\eps
\end{align}
allows us to write the following limiting equations
\begin{equation}
  \Box \W{k}(t,r) = C_{p,k} \frac{\delta^{(k)}(t-r)}{r^p}
\end{equation}
for $k=0,...,n$ which can be solved exactly
\begin{equation}
  \W{k}(t,r) = B_{p,k} \frac{\Theta(t-r)}{r} \left[ \frac{1}{(t-r)^{p+k-2}} - \frac{1}{(t+r)^{p+k-2}} \right]
\end{equation}
with $B_{p,k}:= 2^{p+k-3} C_{p,k} / (p+k-2)$.

The first error term $\h{W}$ comes from the truncation of the expansion \eqref{u0-delta} of $\eps^{-a} \left[ \eps^{-a} u_0(\eps^{-a} t, \eps^{-a} r) \right]^{p}$ into deltas and can be bound by
\begin{equation}
  {\<t+r\>\<t-r\>^{p+n-2}}|\h{W}(t,r)| \leq C
\end{equation}
while the second error term $\w{W}_\eps$ satisfies
\begin{align}
  \Box \w{W}_\eps = \sum_{k=1}^p \begin{pmatrix} p\\ k\end{pmatrix} \eps^{-a} \left[ \eps^{-a} u_0(\eps^{-a} t, \eps^{-a} r) \right]^{p-k} \eps^{[(p-1)+a(p-2)]k} W_\eps^k(t,r).
\end{align}
In \cite{NS-Scaling} we have shown that it can be bound
\begin{equation}
  {\<t+r\>\<t-r\>^{p-2}}|\w{W}_\eps(t,r)| = \O\left({\eps^{(p-1)(1-a)+a[(p-1)^2-2]/p}}\right)
\end{equation}
uniformly for all $t-r>1$. It vanishes in the limit $\eps\ra 0$ for $0<a<p(p-1)/(p+1)$.

It allows to write the asymptotics of $u$
\begin{equation}
\begin{split}
  u(t,r) &= \eps u_0(t,r) + \eps^{p+a(p-1)} W_\eps( \eps^a t, \eps^a r) \\
  &= \eps u_0(t,r) +
  \eps^p \sum_{k=0}^n B_{p,k} \frac{\Theta(t-r)}{r} \left[ \frac{1}{(t-r)^{p+k-2}} - \frac{1}{(t+r)^{p+k-2}} \right]\\
  &+ \O\left(\frac{\eps^p}{\<t+r\>\<t-r\>^{p+n-2}}\right)
  + \O\left(\frac{\eps^{p+\la_0}}{\<t+r\>\<t-r\>^{p-2}}\right)
\end{split}
\end{equation}
which holds uniformly in $t-r>\eps^{-a}$ for $\eps\ra 0$, a given scaling parameter $0<a<p(p-1)/(p+1)$, any nonnegative integer $n$ and $\la_0 := {(p-1)(1-a)+a[(p-1)^2-2]/p}$.

This procedure can be made rigorous as has been demonstrated in \cite{NS-Scaling}.


\begin{acknowledgments}
I would like to thank Piotr Bizo{\'n} and An{\i}l Zengino\u{g}lu for fruitful discussions as well as to Miko{\l}aj Korzy{\'n}ski for his contribution to solving some technical details.

I also kindly acknowledge the support and hospitality of the Institute Mittag-Leffler in Stockholm where parts of this work have been done.
\end{acknowledgments}

\bibliography{QNMs}
\bibliographystyle{unsrt}

\end{document}